\documentclass[conference]{IEEEtran}
\IEEEoverridecommandlockouts
\usepackage{cite}
\usepackage{amsmath,amssymb,amsfonts}
\usepackage{algorithmic}
\usepackage{graphicx}
\usepackage{textcomp}
\usepackage{tikz}
\usepackage{mathtools}
\usepackage{multirow}
\usepackage{arydshln} 
\usetikzlibrary{positioning, arrows.meta, shapes.geometric}

\usepackage{xcolor}
\def\BibTeX{{\rm B\kern-.05em{\sc i\kern-.025em b}\kern-.08em
    T\kern-.1667em\lower.7ex\hbox{E}\kern-.125emX}}
\begin{document}

\newcommand{\inputnum}{3}
\newcommand{\hiddennum}{5}
\newcommand{\outputnum}{1}

\title{Feasibility of short blocklength Reed-Muller codes for physical layer security in real environment}

\author{
Md Munibun Billah, Tyler Sweat, Willie K. Harrison \\
\textit{Department of Electrical and Computer Engineering, Brigham Young University, Provo, Utah, USA} \\
Email: billahm@byu.edu, tysweat0@gmail.com, willie.harrison@byu.edu
}

\maketitle

\begin{abstract}
In this paper, we investigate the application of Reed-Muller (RM) codes for Physical-layer security in a real world wiretap channel scenario. Utilizing software-defined radios (SDRs) in a real indoor environment, we implement a coset coding scheme that leverages the hierarchical structure of RM codes to secure data transmission. The generator matrix of the RM code is used to partition codewords into cosets in the usual way, where each message corresponds to a unique coset, and auxiliary bits select specific codewords within each coset. This approach enables the legitimate receiver (Bob) can decode the transmitted message with minimal information leakage to eavesdropper (Eve) thus  protecting the confidentiality of the communication with the help of coset structure. Mutual information neural estimation (MINE) is used to quantify information leakage and validate the effectiveness of the scheme. Experimental results indicate that RM codes can achieve robust security even in practical environments affected by real-world channel impairments. These findings demonstrate the potential of RM codes as an efficient solution for physical-layer security, particularly for applications that require low latency and short blocklengths.
 \end{abstract}


\section{I\normalfont ntroduction}

Physical-layer security has garnered significant attention as a method for ensuring secure communication in the presence of potential eavesdroppers. This approach leverages the inherent characteristics of communication channels for adding security measures, rather than relying solely on traditional cryptographic techniques. The wiretap channel model, first introduced by Wyner in 1975, provides a theoretical foundation for this type of security by examining how information can be securely transmitted even when an adversary has access to the communication channel. In Wyner’s model, a legitimate receiver (Bob) and an eavesdropper (Eve) receive different versions of the transmitted signal due to the varying conditions of the communication channels, and secrecy is achieved when the main channel (Alice to Bob) is more reliable than the wiretap channel (Alice to Eve) .

This work explores the practical application of wiretap codes, particularly Reed-Muller codes, in achieving Physical-layer security. By implementing these codes on software-defined radios (SDRs), we assess the performance of small blocklength coset codes under real world conditions. Specifically, we evaluate the bit error rate (BER) and information leakage for coded and uncoded transmissions, across varying distances between the transmitter and the eavesdropper, demonstrating regions where the legitimate receiver gains a secrecy advantage. To calculate the information leakage, we employ Mutual Information Neural Estimation (MINE) and analyze the effectiveness of the wiretap coding scheme in an indoor environment.

Information leakage the mutual information between the confidential message and the version of message received by Eve. Mutual information in information theory quantifies the amount of information one random variable contains about another. Several techniques have been developed to calculate mutual information, each with its own advantages and applications. Analytical calculation is possible when the joint probability distribution is known, allowing mutual information to be computed directly using integrals or sums \cite{cover}. However, in practical scenarios where it is difficult to know the distributions, several methods have been such as density-based estimation, bayesian approach, K-nearest neighbors method developed that require empirical estimation \cite{mutual_inf},\cite{density}, \cite{mutual_inf}, \cite{bayesian}. \cite{mutual_inf} can suffer from bias due to bin size selection. Kernel density estimation requires careful selection of kernel parameters \cite{density}. The K-nearest neighbors method can be computationally intensive \cite{mutual_inf}. More recently, Mutual Information Neural Estimation (MINE) has been introduced, leveraging deep learning to estimate mutual information by optimizing neural networks that bound the mutual information from below \cite{MINE}. MINE is especially powerful for high-dimensional and complex distributions, though it requires careful tuning of the neural network architecture and is computationally expensive. This study uses the MINE to calculate information leakage due to high dimensionality of the applied coset codes.

Even though the physical-layer security specially coset coding has seen significant theoretical advancement there is a scarcity of studies exploring practical implementation resulting a substantial gap between theory and real-world applications. Most of the existing work has explored theoretical aspects of coset coding focusing on channel conditions and coding strategies. Several works propose frameworks for achieving secrecy capacity and demonstrate coset coding's effectiveness in theoretical limits. However, the translation of these theoretical work to real world is rare because of the dynamic nature of coset based schemes and noisy real world channels \cite{yang}, \cite{experiment_modular}. The limited experimental works existing in the literature often make use of the knowledge about the channel \cite{experiemntal_lattice}, \cite{experi_harrison}. This paper investigates the feasibility of Coset based wiretap coding using Reed-Muller codes in real world environment. Transmitting data using software defined radios (SDRs), this study evaluates coset based codes capability to provide physical-layer security using deep learning based mutual information calculation. 

The remainder of this paper is organized as follows. Section II discusses the theoretical background, including the wiretap channel model and the Reed-Muller code used in our experiments. Section III details the experimental setup, including the environment, procedure, and the SDR configurations. Section IV presents the results, including the performance of BER, mutual information analysis, and a heat map of the results of the collected data. Finally, Section V concludes the paper with a discussion of the findings and future research direction.

\section{W\normalfont iretap C\normalfont ode}
\label{sec:wiretap}
\noindent

\begin{figure}[!ht]
\begin{center}
    	\includegraphics[scale=.4,clip, trim=1.15cm 13.5cm 1.2cm 7.5cm ]{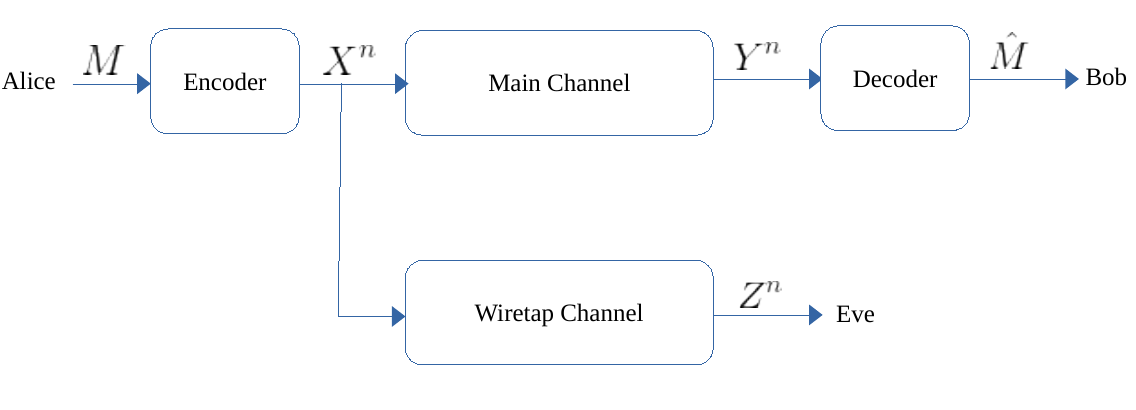}
	\caption{The wiretap channel.}
	\label{fig:wiretapchannel}
\end{center}
\end{figure}

\noindent 
The wiretap channel was introduced by Wyner in his seminal paper \cite{Wyner} and a version of it is shown in Figure \ref{fig:wiretapchannel}. Here Alice is sending a confidential message $M$ to Bob by encoding it to $X^n$ and transmitting it over the main channel. Bob sees the main channel output $Y^n$ and estimates the confidential message $\hat{M}$. There is an eavesdropper Eve who also receives a version of the transmission $Z^n$ over the wiretap channel. Here all the capital letters denote random variables and lowercase letters are the realization of corresponding random variables. It is assumed that Eve has perfect knowledge of the encoder and decoder employed by Alice and Bob. One of the consequences of Wyner's result is that to achieve secrecy the wiretap channel has to be noisier than the main channel \cite{positive_secrecy}. 

\subsection{ Secrecy Metric}
To provide an information-theoretic security argument, secrecy is quantified by the mutual information between confidential message $M$ and wiretap channel output $Z^n$, given by $I(M; Z^n)$.

Among the defined criteria for information-theoretic secrecy, strong secrecy \cite{strong_secrecy} and weak secrecy \cite{Wyner} are the prominent ones in literature. Strong secrecy is achieved if 
\begin{equation}
    \lim_{n \to \infty } I(M; Z^n)=0,
\end{equation}
and weak secrecy is achieved if
\begin{equation}
    \lim_{n \to \infty } \frac{1}{n} I(M; Z^n)=0.
\end{equation}
Both of these definitions require the calculation of $I(M; Z^n)$. They also imply statistical independence between $M$ and $Z^n$ as $n$ approaches infinity. While strong secrecy is more desirable weak secrecy is generally easier to achieve.  

\subsection{Different Wiretap Coding}
Various coding strategies have been developed to enhance security in wiretap channels \cite{harrison_review}.  Polar codes \cite{polar_main}, introduced by Arikan, achieve secrecy in degraded wiretap settings by leveraging channel polarization. Polar codes limits information leakage by assigning bad channels to  the eavesdropper, thus limiting Eve’s ability to decode the transmitted message. \cite{polar}. By adjusting degree distributions and puncturing strategies, Low-Density Parity-Check (LDPC) codes have also been applied to wiretap channels due to their strong error-correcting capabilities and ubiquitousness \cite{LDPC_application}. \cite{munibun}, introduces an approach to enhance LDPC codes for wiretap channels by injecting artificial noise, effectively reducing information leakage to the eavesdropper in Gaussian channels. Lattice codes are particularly studied for Gaussian wiretap channels, where high-dimensional or continuous-valued signals are involved. It secrecy by mapping the eavesdropper’s observations ambiguously within the lattice, making it highly effective for non-binary wiretap channels \cite{lattice}.  RM codes \cite{RM_main} are also a viable choice for coset coding due to their structured nature, which allows for efficient partitioning into cosets to further increase uncertainty for the eavesdropper [5]. Notably, RM codes can achieve capacity in binary erasure channels (BEC) \cite{RM_capacity}, making them effective in scenarios where secrecy and reliability are needed in tandem.

\section{C\normalfont oset C\normalfont oding}
Coset coding is a technique used in wiretap coding to enhance the reliability and security of transmitted data, particularly in scenarios where error correction and information security are crucial \cite{harrison_review}. Coset coding involves the partitioning of a vector space into disjoint subsets, known as cosets, relative to a linear code. Each coset is formed by adding a fixed vector, called the coset leader, to all codewords in a given linear code.

\subsection{Coset Coding: Encoder and Decoder}

Let $C$ be an $(n, n - k)$ binary linear block code, and $C_0 = C, C_1, C_2, \dots, C_{2^k - 1}$ be the cosets of $C$. If $G$ is an $(n - k) \times n$ generator matrix of $C$ and $H$ is a $k \times n$ parity-check matrix of $C$, then
\[
G' = \begin{bmatrix} G \\ G^* \end{bmatrix}
\]
is the generator matrix for a coset-style secrecy code, where $G^*$ is comprised of $k$ basis vectors of $\mathbb{F}_2^n$ that are not in $C$. The coset-style secrecy encoding is then done by concatenating an $(n - k)$-bit uniformly distributed binary auxiliary message $\Tilde{M}$ with a uniformly distributed $k$-bit message $M$ and codewords are computed as 
\[
x^n = \begin{bmatrix} m & \Tilde{m} \end{bmatrix} G'.
\]
 The message $M$ chooses the coset, and the auxiliary message $\Tilde{M}$ chooses a specific codeword from the coset uniformly at random. Thus, the encoder is a one-to-many mapping from message to codewords.

For example, consider the secrecy codebook given in Table I. Here, $C = C_0$ is a $(4, 2)$ linear code with generator matrix
\[
G = \begin{bmatrix} 0 & 0 & 1 & 1 \\ 1 & 1 & 0 & 1 \end{bmatrix},
\]
and coset style secrecy code generator matrix is
\[
G' = \begin{bmatrix} 0 & 0 & 1 & 1 \\ 1 & 1 & 0 & 1 \\ 0 & 1 & 1 & 1 \\ 1 & 1 & 0 & 0 \end{bmatrix}.
\]
The generator matrix gives the codebook described as in the table \ref{table:codebook}.

\begin{table}[h!]
    \centering
    \caption{Code table for a best $n = 4$, $k = 2$ coset-style secrecy code.}
    \begin{tabular}{|c|c|c|c|c|}
        \hline
        \textbf{Coset} & $\Tilde{m}=00$ & $\Tilde{m}=01$ & $\Tilde{m}=10$ & $\Tilde{m}=11$ \\ \hline
        $C_0 \; (m = 00)$ & 0000 & 1101 & 0011 & 1110 \\ \hline
        $C_1 \; (m = 01)$ & 1100 & 0001 & 1111 & 0010 \\ \hline
        $C_2 \; (m = 10)$ & 0111 & 1010 & 0100 & 1001 \\ \hline
        $C_3 \; (m = 11)$ & 1011 & 0110 & 1000 & 0101 \\ \hline
    \end{tabular}
    \label{table:codebook}
\end{table}

To decode the receiver calculates the syndrome to obtain
\begin{equation}
\begin{split}
    s^k &= y^n H^T\\ 
        &= x^n H^T \\
        &= \begin{bmatrix} m^k & \Tilde{m}^{n-k} \end{bmatrix} 
        \begin{bmatrix} G^T \\ G^{*T} \end{bmatrix} H^T \\
        &= m^k (G^T H^T) + \Tilde{m}^{n-k} (G^{*T} H^T) \\
        &= m^k,
\end{split}
\end{equation}

since $G^T H^T = I_k$ and $G^{*T} H^T = 0$ by definition, where $I_k$ is the $k \times k$ identity matrix.

\subsection{Reed Muller Code}

The Reed-Muller code \(\text{RM}(r, m)\) is a linear block code defined by two parameters: the order \(r\) and the length \(m\). The code \(\text{RM}(r, m)\) is a subspace of the vector space \(\mathbb{F}_2^{2^m}\) over the binary field \(\mathbb{F}_2\).

The generator matrix \(G\) of \(\text{RM}(r, m)\) is constructed from all monomials of degree at most \(r\) in \(m\) variables. If we denote the variables by \(x_1, x_2, \dots, x_m\), then the monomials are of the form:

\[
x_1^{a_1} x_2^{a_2} \cdots x_m^{a_m} \quad \text{where} \quad a_i \in \{0, 1\} \quad \text{and} \quad \sum_{i=1}^{m} a_i \leq r
\]

 The length \(n\) of the codewords in \(\text{RM}(r, m)\) is given by:
  \[
  n = 2^m
  \]

The dimension \(k\) of \(\text{RM}(r, m)\) is the number of monomials of degree at most \(r\), which is calculated as:
  \[
  k = \sum_{i=0}^{r} \binom{m}{i}
  \]

Each codeword in \(\text{RM}(r, m)\) corresponds to a polynomial in \(m\) variables where the coefficients of the polynomial are binary (i.e., 0 or 1). The codeword associated with a polynomial \(f(x_1, x_2, \dots, x_m)\) is the evaluation of \(f\) at all possible binary inputs:

\[
c = \left(f(0,0,\dots,0), f(0,0,\dots,1), \dots, f(1,1,\dots,1)\right)
\]

\section{Coset Coding with Reed-Muller Codes}

In this study, Reed-Muller (RM) codes are utilized to implement a coset coding scheme for secure communication over a wiretap channel due to their structured design, which enables efficient partitioning into cosets. We employ a full-rank, square generator matrix $ G^{'}$ of dimension $ n \times n $ derived from the RM code, which ensures that each coset can uniquely represent a message while providing additional randomness within each coset.

\subsection{Reed-Muller Code Structure and Generator Matrix}

The generator matrix \( G^, \) used in this setup is constructed so that:
\begin{itemize}
    \item The first \( n/2 \) columns correspond to the message bits \( m \), which define the coset of the code.
    \item The remaining \( n/2 \) columns correspond to the auxiliary bits \( \Tilde{m} \), generated uniformly at random to select specific codewords within each coset.
\end{itemize}
The Reed-Muller structure ensures that each row of $ G^{'}$,  is linearly independent, enabling unique coset representations for each possible message $m$.

In this coset coding approach, each message is mapped to a unique coset of the Reed-Muller code. The encoding process for Alice calculates a codeword \( x^n \) as follows:

\[
x^n = \begin{bmatrix} m & \Tilde{m} \end{bmatrix} G^{'} = m G + \Tilde{m} G^{*}
\]

where:
\begin{itemize}
    \item \( m G \): This part of the codeword determines the coset of the Reed-Muller code. Each unique \( m \) selects a distinct coset within the code, leveraging the structured redundancy of Reed-Muller codes. This ensures that each transmitted message belongs to a different subset of codewords.
    \item \( \Tilde{m} G^* \): The auxiliary bits \( \Tilde{m} \) are chosen uniformly at random and used to select a specific codeword within the chosen coset. This random selection within each coset enhances security by ensuring multiple possible codewords for each message.
\end{itemize}
 This setup allows each coset to represent a different message while adding randomness through the auxiliary bits.

For example, in this study the Reed-Muller code \(\text{RM}(3, 3)\) is used and the code generated by the monomials of degree at most 3 in three variables \(x_1\), \(x_2\), and \(x_3\). The generator matrix \(G\) is

\begin{equation}   
G' =\begin{bmatrix} G \\ 
G^* \end{bmatrix} =
\begin{bmatrix}
1 & 1 & 1 & 1 & 1 & 1 & 1 & 1 \\
0 & 0 & 0 & 0 & 1 & 1 & 1 & 1 \\
0 & 0 & 1 & 1 & 0 & 0 & 1 & 1 \\
0 & 1 & 0 & 1 & 0 & 1 & 0 & 1 \\
\hdashline
0 & 0 & 0 & 0 & 0 & 0 & 1 & 1 \\
0 & 0 & 0 & 0 & 0 & 1 & 0 & 1 \\
0 & 0 & 0 & 1 & 0 & 0 & 0 & 1 \\
0 & 0 & 0 & 0 & 0 & 0 & 0 & 1
\end{bmatrix}
\end{equation} 
Here the first $4$ rows of $G'$ corresponds to the matrix $G$ and last $4$ corresponds to the matrix $G^*$.The use of a full-rank, square generator matrix \( G^* \) derived from the Reed-Muller code ensures that each possible message maps to a unique coset, thereby maximizing the uncertainty for an eavesdropper. This approach leverages both the redundancy and structure of Reed-Muller codes, enhancing security by making the transmitted codewords indistinguishable to unauthorized listeners, as each message can correspond to several potential codewords.
This coset coding method, based on Reed-Muller codes, effectively uses both coset selection and intra-coset randomization to provide a secure encoding scheme in the wiretap channel model.

\section{M\normalfont utual I\normalfont nformation N\normalfont eural E\normalfont stimator}
\label{sec:Mine}

\begin{figure}[!ht]
\end{figure}
To calculate the leakage $I(M; Z^n)$, this work uses the mutual information neural estimator (MINE). MINE uses the Donsker-Varadhan representation of the Kullback-Leibler divergence
\begin{equation} \label{eq:Donsker}
    D_{KL}(P||Q)=\sup_{F:\Omega\rightarrow \mathbb{R}} \mathbb{E}_P[f(X, Y)]-\log(\mathbb{E}_Q[e^{f(X, Y)}]),
\end{equation}
where the supremum is taken over all functions $f$ such that the expectations are finite for random variables $X$ and $Y$. The relation between mutual information and Kullback-Leibler divergence is given by 
\begin{equation}
    I(X; Y)=D_{KL}(P_{XY}||P_{X}P_{Y}).
\end{equation}
In \cite{MINE}, the authors proposed to choose $\mathcal{F}$ to be a set of functions $T_\theta:\mathcal{X}\times \mathcal{Y}\rightarrow \mathbb{R}$ parameterized by deep neural network $\theta \in \Theta$. 
Since (\ref{eq:Donsker}) is a lower bound for Kullback-Leibler divergence with equality with an optimal function choice, 
\begin{equation}
    I(X; Y) \geq I_{\Theta}(X; Y), 
\end{equation}
where
\begin{equation} \label{eq:est}
    I_{\Theta}(X; Y)= \sup_{\theta \in \Theta} \mathbb{E}_{p(x, y)}[T_\theta(X, Y)]-\log(\mathbb{E}_{p(x)p(y)}[e^{T_\theta(X, Y)}]).
\end{equation}
Now, if we identify $X$ as a confidential message $M$ and $Y$ as the wiretap channel output $Z^n$, then $p(M, Z^n)$ is the joint probability distribution, and $p_{M}$, $p_{Z^n}$ are the respective marginals of $(M, Z^n)$. Then (\ref{eq:est}), can be written as 
\begin{equation} \label{eq:esti}
\begin{multlined}
    I_{\Theta}(M; Z^n)= \sup_{\theta \in \Theta} \mathbb{E}_{p(M, Z^n)}[T_\theta(M, Z^n)]\\
    -\log(\mathbb{E}_{p(M)p(Z^n)}[e^{T_\theta(M, Z^n)}]). 
\end{multlined}
\end{equation}
Since in practice, the true distribution  $p(M, Z^n)$ is unknown, we can't use $I_{\Theta}(M; Z^n)$ to estimate $I(M; Z^n)$. Rather we can estimate expectations presented in the equation (\ref{eq:esti})  using the samples of joint and marginal distributions, by rewriting $I_{\Theta}(M; Z^n)$
\begin{equation} \label{eq:MINE}
    \hat{I}(M; Z^n):=\frac{1}{l}\sum_{i=1}^{l} [T_\theta (m_i,z_i^n)]-\log \frac{1}{l} \sum_{i=1}^{l}[e^{T_\theta(\overline{m}_i,\overline{z}_i^n)}],
\end{equation}
where $l$ is the number of samples. In (\ref{eq:MINE}), $l$ samples of the joint distribution are generated by producing uniformly distributed confidential messages $m$, and from wiretap channel outputs $z^n$. The term $(\overline{m}_i,\overline{z}_i^n)$ represents samples generated from marginal distributions. The network used in this work has five fully connected hidden layers with each layer consisting of $500$ nodes and Relu activation functions. The input layer has $k+n$ neurons. During the training, ten thousand messages with a batch size of $1000$ were used. We also used the Adam optimizer \cite{adam} with a learning rate of $10^{-7}$ and $2.5\times10^5$ epochs.

\begin{figure*}[!ht]
\centering
    \includegraphics[width=1\linewidth]{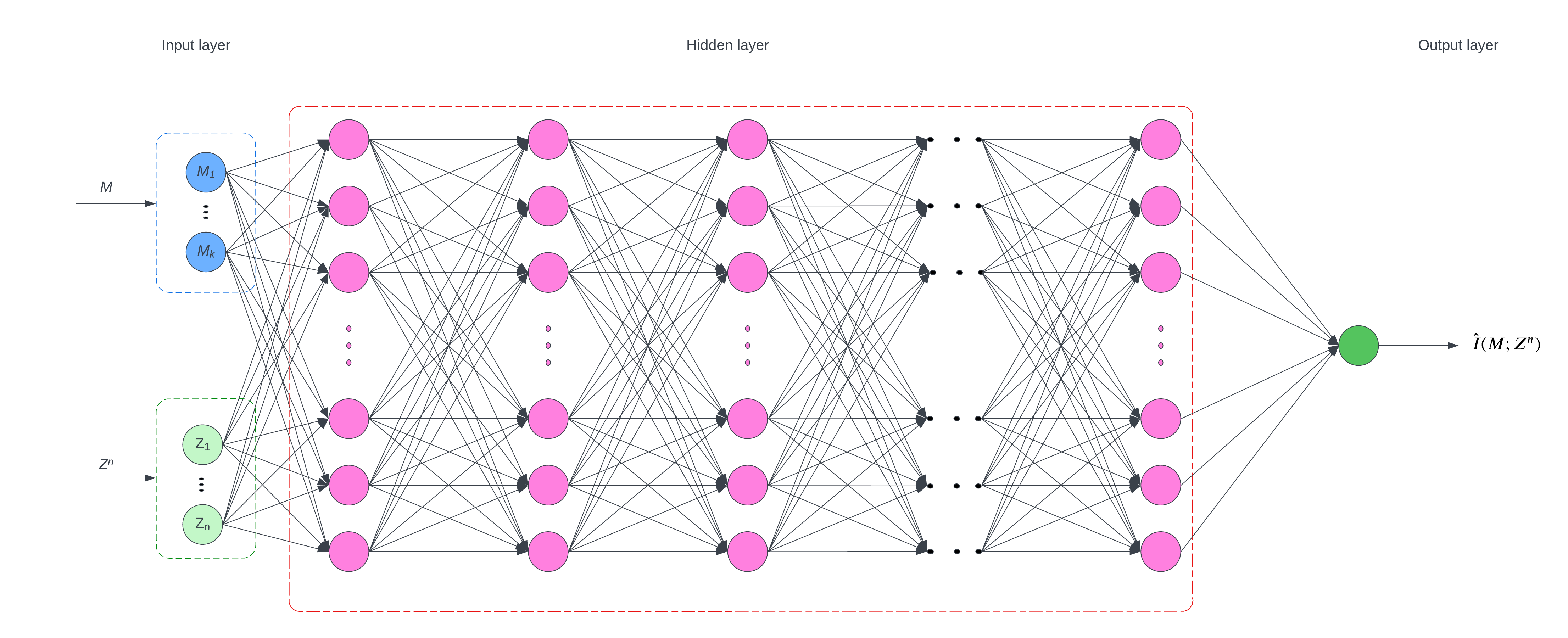}
    \caption{Mutual Information Neural Estimator.}
    \label{fig:mine}
\end{figure*}

\section{E\normalfont quivocation C\normalfont alculation}

In this study, the equivocation \( E \) represents the uncertainty that an eavesdropper (Eve) has about the transmitted message \( M \) after observing the signal \( Z^n \). The equivocation is defined as the conditional entropy \( H(M | Z^n) \), which can be calculated by determining the mutual information between \( M \) and \( Z^n \) and then subtracting it from the entropy of \( M \).

The equivocation \( E \) is given by:
\[
E = H(M | Z^n) = H(M) - I(M; Z^n),
\]
where:
\begin{itemize}
    \item \( H(M) \) is the entropy of the message \( M \), representing the initial uncertainty about \( M \) before any observation by Eve. Since \( H(M) \) depends only on the distribution of \( M \), it is constant for a given message distribution.
    \item \( I(M; Z^n) \) is the mutual information between \( M \) and \( Z^n \), which quantifies the amount of information that Eve’s observations \( Z^n \) provide about \( M \).
\end{itemize}

In this study, we calculate the mutual information \( I(M; Z^n) \) using Mutual Information Neural Estimation (MINE), a data-driven technique that approximates mutual information through neural networks. MINE enables flexible and adaptive estimation of mutual information, accounting for complex dependencies between \( M \) and \( Z^n \) that might not be captured by traditional methods. By applying MINE, we obtain an accurate measure of the information Eve gains about the message.

To compute the equivocation \( E \), we use:
\[
E = H(M) - I(M; Z^n),
\]
where \( H(M) \) is constant and \( I(M; Z^n) \) is estimated through MINE.

This approach to calculating equivocation, with mutual information estimated via MINE, ensures that we capture any information leakage that Eve gains through her observations. High equivocation indicates strong security, as it implies that Eve has high uncertainty about the original message \( M \) despite observing \( Z^n \).

\section{E\normalfont xperimental S\normalfont etup}

\subsection{Environment}
The experiments were conducted in an indoor setting, specifically in the step-down lounge study area of the Clyde Building on the campus of Brigham Young University. This area features an open space layout with three narrow pillars down the center, and the north, west, and south walls are composed of windows and cinder blocks. Data were collected over a span of multiple hours
\subsection{Channel Model}

The study was conducted in a real indoor environment, which inherently introduces channel effects such as multipath propagation, path loss, shadowing, and background noise. Although these effects were not explicitly modeled, they naturally influence the signal characteristics in ways that impact both reliability and security.

\begin{itemize}
    \item \textbf{Multipath Propagation and Fading}: Signals reflect off walls and objects, causing multipath propagation, which can lead to fading effects. These include constructive and destructive interference, resulting in fluctuations in signal strength.
    
    \item \textbf{Path Loss and Shadowing}: Signal strength decreases over distance, with additional attenuation caused by obstructions like walls and furniture. Random fluctuations in received signal power, due to these obstacles, also play a role.
    
    \item \textbf{Noise and Interference}: Background noise from electronic devices and thermal noise contribute to the overall signal quality. This interference affects the clarity of the received signal.

    \item \textbf{Implications for Security}: The inherent variations in an indoor channel create differences between the main channel (Alice to Bob) and any potential eavesdropper’s channel (Alice to Eve), providing a level of natural security at the Physical-layer.
\end{itemize}

\subsection{SDR Configuration and Setup}
Two ADALM-PLUTO software-defined radios (SDRs) with stock antennas were used for secure data transmission and reception. One SDR was configured as the transmitter (Alice) and the other as the receiver (Bob), both set to a transmission frequency of \( 915 \ \text{MHz} \), within the industrial, scientific, and medical (ISM) band, making it suitable for indoor environments.

\subsection{Transmission Parameters}
\begin{itemize}
    \item \textbf{Sample Rate}: The sample rate for both SDRs was set to \( 1 \ \text{MHz} \), providing high-resolution signal sampling.
    \item \textbf{Samples per Symbol}: Each symbol was represented by \( 8 \) samples, balancing signal resolution and processing efficiency.
\end{itemize}

\subsection{Loop Filter (LF) Configuration}
The loop filter (LF) was configured to stabilize the phase and frequency of the transmitted signal, with the following parameters:
\begin{itemize}
    \item \( \text{LF $B_nT$} = 0.001 \): Loop bandwidth to symbol rate product.
    \item \( \text{LF $\zeta$} = 0.7071 \): Damping factor set to achieve critical damping.
\end{itemize}

\subsection{Phase and Timing Error Detector Configuration}
\begin{itemize}
    \item \textbf{Phase Error Detector (PED)}: Configured to correct phase errors with the parameters:
        \begin{itemize}
            \item \( \text{PED $K_0$} = 1 \): Initial gain.
            \item \( \text{PED $K_1$} = 0.1479 \): Proportional gain.
            \item \( \text{PED $K_2$} = 0.0059 \): Integral gain.
        \end{itemize}
    \item \textbf{Timing Error Detector (TED)}: Configured to ensure symbol timing synchronization with:
        \begin{itemize}
            \item \( \text{TED $K_p$} = 2.8 \): Proportional gain.
            \item \( \text{TED $K_0$} = -1 \): Initial gain.
            \item \( \text{TED $K_1$} = -9.51 \times 10^{-4} \): Fine adjustment coefficient.
            \item \( \text{TED $K_2$} = -1.27 \times 10^{-6} \): Integral gain.
        \end{itemize}
\end{itemize}

\subsection{Procedure}
\begin{figure*}[!ht]
    \includegraphics[width=\linewidth]{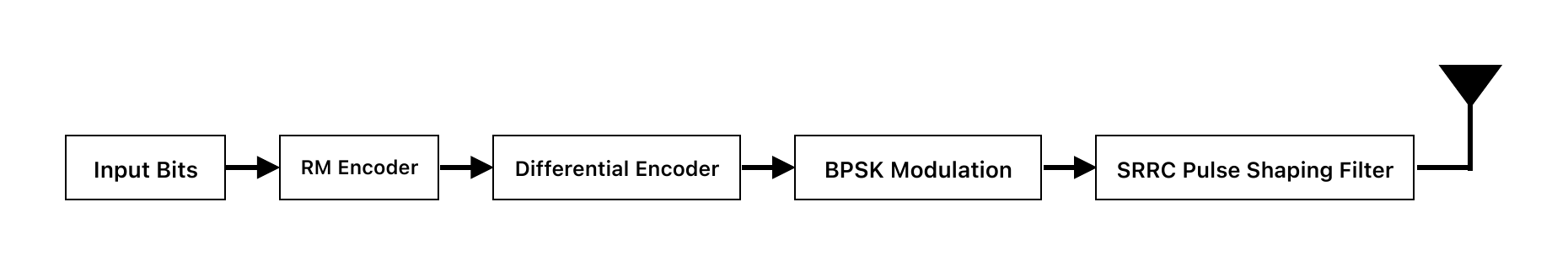}
    \caption{Transmitter Block Diagram with Reed-Muller Wiretap Coding}
    \label{fig:tx_block}
\end{figure*}

\begin{figure*}[!ht]
    \centering
    \includegraphics[width=\linewidth]{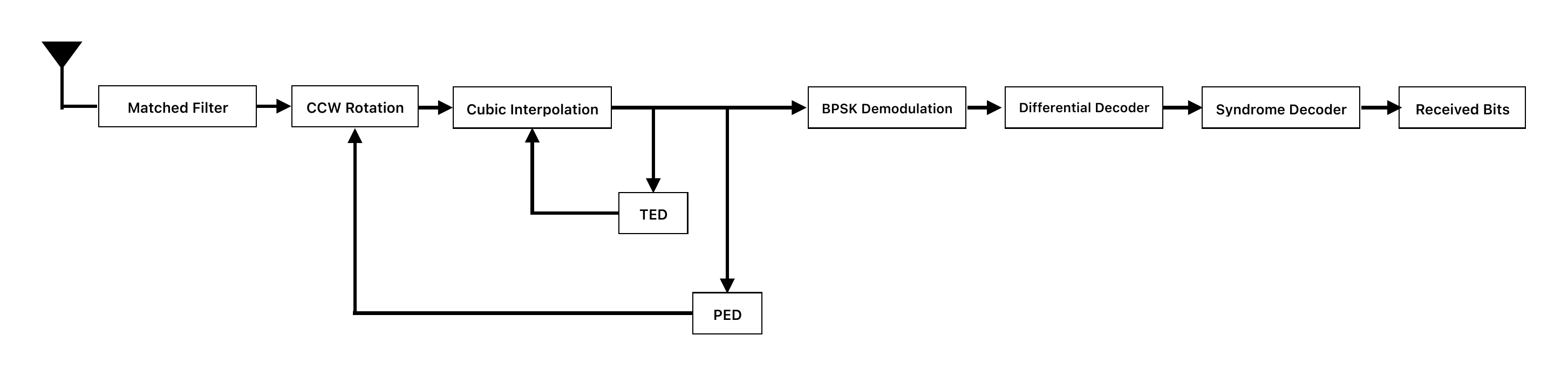}
    \caption{Receiver block diagram.}
    \label{fig:rx_block}
\end{figure*}

A BPSK transmitter/receiver pair with differential encoding \cite{rice} was implemented according to block diagrams as shown in Figures \ref{fig:tx_block} and \ref{fig:rx_block}. Two ADALM-PLUTO SDR modules were used in the experiment, each equipped with the stock antenna.
The transmitter was located at the origin of the map shown and constantly transmitted a known bitstream. The receiver was moved throughout the room at intervals of 3 feet in a grid around the origin. At each test point, samples were received from the Pluto SDR and the receiver attempted to recover the bits sent. The resulting uncoded and coded bit error rate (BER) were calculated and recorded along with the recovered bit stream. 

For each test point, the receiver processed the received samples and attempted to recover the transmitted bits. The uncoded and coded bit error rate (BER) were calculated and recorded for each position along with the recovered bitstream.

We used the Reed-Muller code \(\text{RM}(4,4)\) to encode message bits combined with auxiliary random bits, producing coded bits for transmission in the wiretap channel. A total of \( n = 8000 \) message bits were generated with an alternating pattern, where even-indexed bits were set to 1 and odd-indexed bits to 0. Using the full-rank \( 16 \times 16 \) generator matrix \( G^{'} \) of the Reed-Muller code, each group of 8 message bits \( m \) was paired with an auxiliary 8-bit vector$\Tilde{m}$, randomly generated to add intra-coset randomness. The combined 16-bit vector \( x = [m, \Tilde{m}] \) was then encoded using the generator matrix $G^{'}$, ensuring each message mapped to a specific coset while \( \Tilde{m} \) selected a codeword within the coset. The resulting \( 16000 \) coded bits were stored for transmission, with the encoded data saved  for further analysis and experimental evaluation.

\section{R\normalfont esults}
\begin{figure}[!ht]
    \includegraphics[width=1\linewidth]{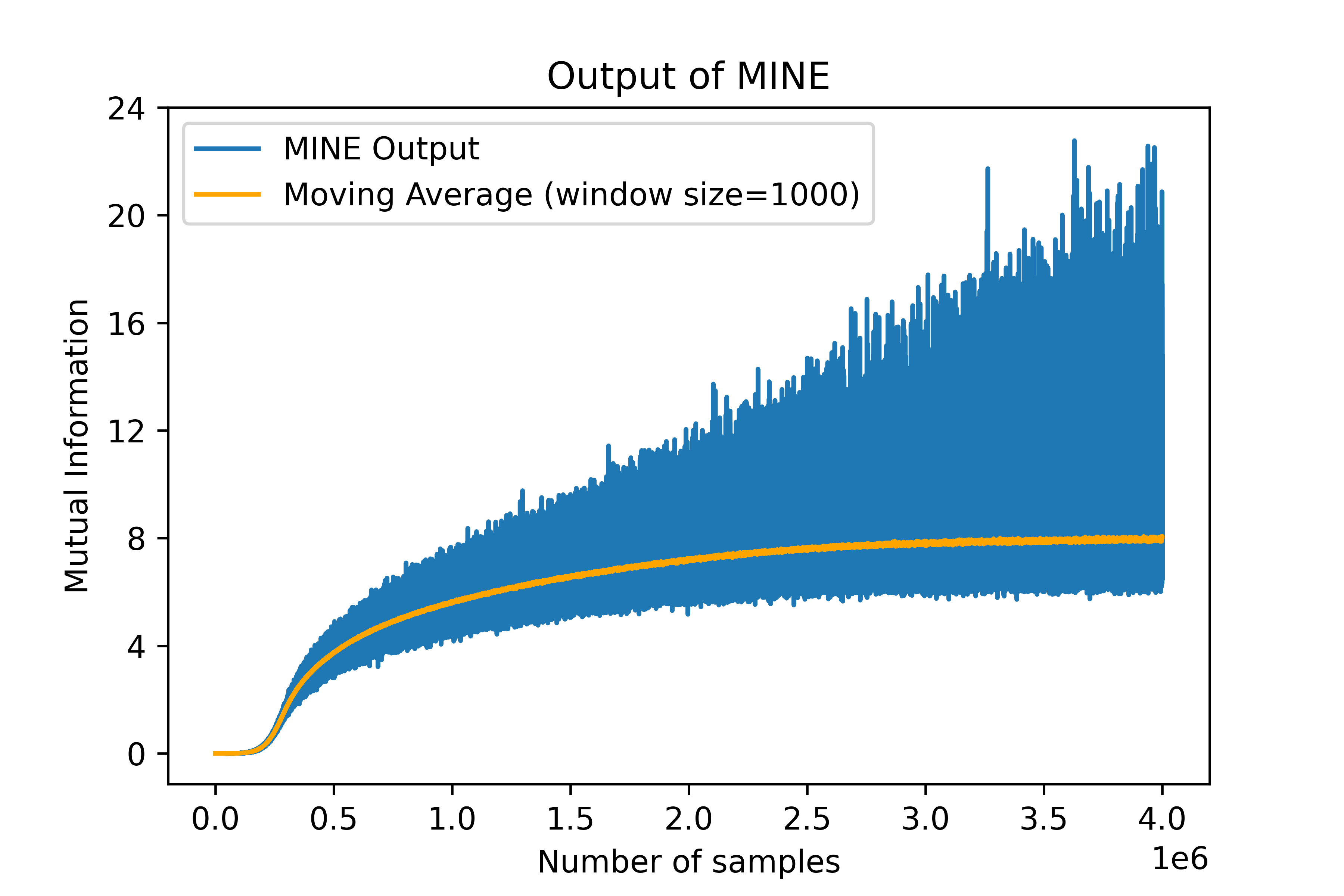}
    \caption{Output of MINE showing mutual information between confidential message and Eaves received message for RM(4,4) coset style code in the presence of no noise.}
    \label{fig:mine_converge}
\end{figure}

The output of Mutual Information Neural Estimation (MINE), shown in Figure~\ref{fig:mine_converge}, provides an estimate of the mutual information between the transmitted and received data. However, as seen in the figure, the MINE output exhibits a significant level of noise, which makes it challenging to interpret the convergence behavior directly. This noise is typical in MINE outputs, especially when estimating mutual information over a large number of samples, as neural estimators can introduce variability due to their stochastic nature. In this study we found that the required number and size of hidden layers and number of epoch to converge increases rapidly with the blocklength of the code. The design configuration and training requirements of MINE for this study has been provided in table \ref{table:rm_codes}. The activation functions used for all network is Relu. Weight initialization for each Layer is $\mathcal{N}(0, 0.02)$ wit zero bias to ensure a stable starting points.

\begin{table*}[h!]
    \centering
    \caption{Network Configuration and Training Requirements for Reed-Muller Codes}
    \resizebox{\textwidth}{!}{%
    \begin{tabular}{|c|c|c|c|}
        \hline
        \textbf{Parameter}           & \textbf{RM(2,2)}   & \textbf{RM(3,3)}   & \textbf{RM(4,4)}   \\ \hline
        \textbf{Hidden Layers}       & 3                  & 4                  & 6                  \\ \hline
        \textbf{Neurons per Layer}   & 50                 & 200                & 500                \\ \hline
        \textbf{Epochs}              & $2 \times 10^5$    & $5 \times 10^5$    & $4 \times 10^6$    \\ \hline
        \textbf{Batch Size}          & 100                & 400                & 1000               \\ \hline
        \textbf{Moving Average Window size}      & 100                & 300                & 1000               \\ \hline
        \textbf{Activation Function} & ReLU               & ReLU               & ReLU               \\ \hline
        \textbf{Initialization}      & Weights: $\mathcal{N}(0, 0.02)$, Bias: 0 & Weights: $\mathcal{N}(0, 0.02)$, Bias: 0 & Weights: $\mathcal{N}(0, 0.02)$, Bias: 0 \\ \hline
    \end{tabular}%
    }
    \label{table:rm_codes}
\end{table*}

To address this, a moving average  with a window size given in table \ref{table:rm_codes} for each codes was applied to smooth the MINE output. The moving average effectively reduces the noise, making it easier to observe the underlying trend in the mutual information estimate. As illustrated by the orange line in Figure~\ref{fig:mine_converge}, the smoothed output shows a clear convergence towards a stable value.

\begin{figure}[!ht]
\centering
    \includegraphics[width=1\linewidth]{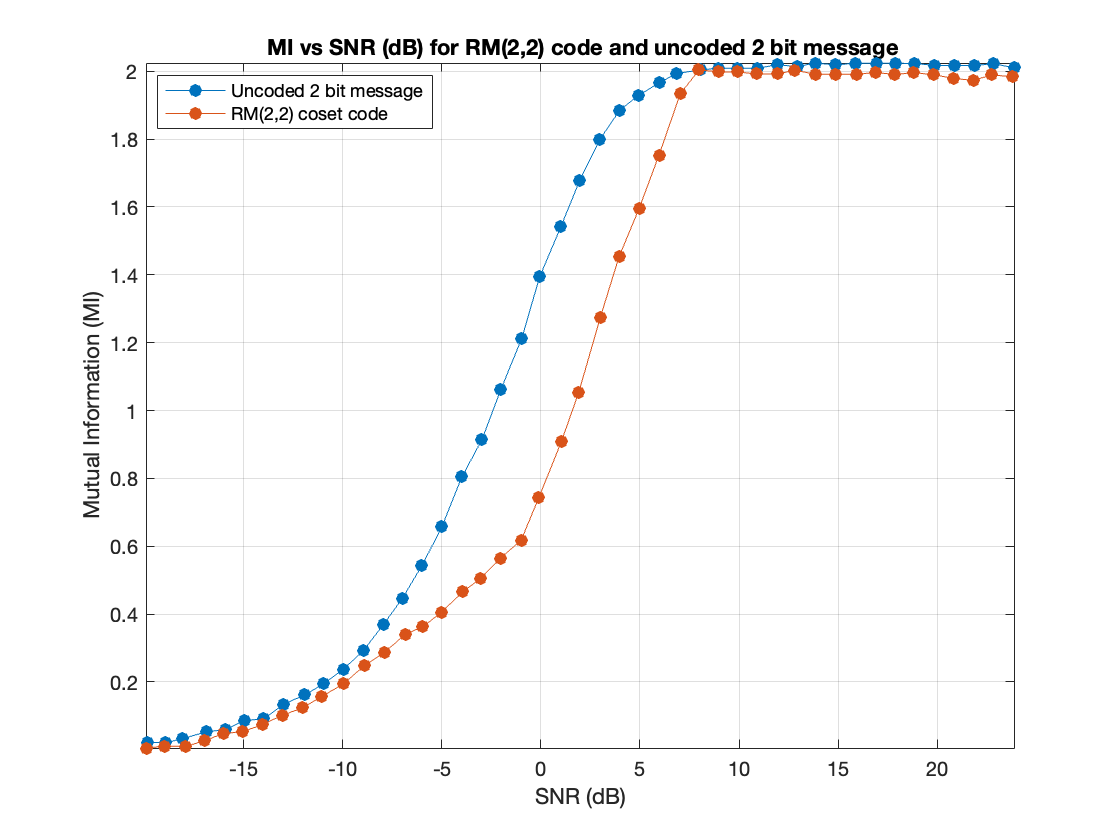}
    \caption{Comparison of Mutual Information Between RM(2,2) and uncoded 2 bit message.}
    \label{fig:rm_22}
\end{figure}

\begin{figure}[!ht]
\centering
    \includegraphics[width=1\linewidth]{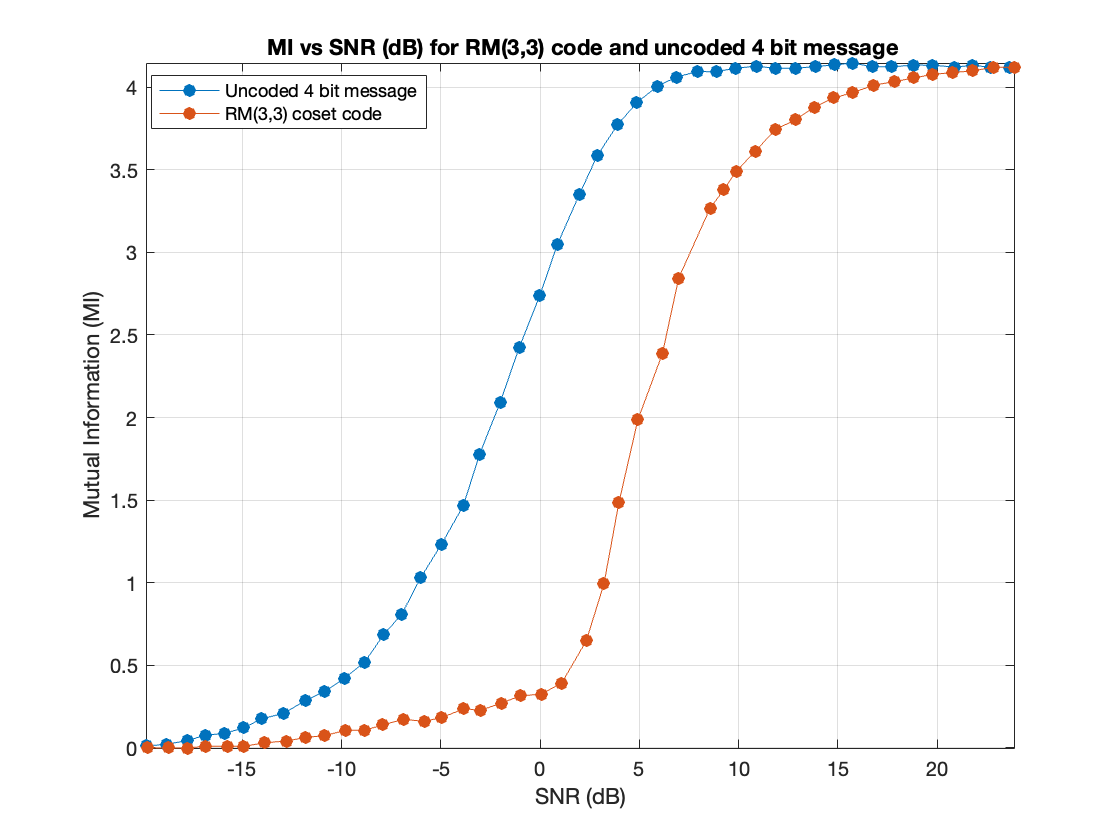}
    \caption{Comparison of Mutual Information Between RM(3,3) and uncoded 4 bit message.}
    \label{fig:rm_33}
\end{figure}

The result of the moving average converges close to the expected mutual information for the Reed-Muller \(\text{RM}(4,4)\) code, which is 8 bits. This is consistent with the theoretical capacity for \(\text{RM}(4,4)\), as it has 16 codewords with 8 information bits, providing all the information are retained in the received bits by Eve in the presence of very low to no noise in Eavesdropper channel.
Figures~\ref{fig:rm_22} and \ref{fig:rm_33} illustrate the mutual information (MI) versus signal-to-noise ratio (SNR) for RM wiretap codes (2,2) and (3,3) alongside their respective uncoded messages of 2 bits and 4 bits. These plots reveal the effectiveness of RM coding in reducing information leakage to an eavesdropper (Eve) across different SNR values.

In Figure~\ref{fig:rm_22}, the mutual information for the RM (2,2) wiretap code grows gradually as the SNR increases, reaching a saturation level  the maximum of 2 bits, which corresponds to the coded message’s information content. Conversely, the uncoded 2-bit message reaches full mutual information (2 bits) more rapidly, indicating that, as SNR improves, Eve can fully decode the uncoded message, resulting in complete information leakage. The RM (2,2) code, however, limits Eve’s information gain even at  SNR values from $-8dB$ to $7dB$, thereby enhancing security.

Similarly, Figure~\ref{fig:rm_33} compares the RM (3,3) wiretap code with an uncoded 4-bit message. Here, the mutual information for the RM (3,3) coded message increases more slowly with SNR and saturates at 4 bits. In contrast, the uncoded 4-bit message quickly reaches a mutual information level of 4 bits as SNR increases, indicating that Eve could decode the message entirely at high SNR. This code provides secrecy from $-11dB$ to $13dB$.

These results demonstrate that Reed-Muller wiretap codes effectively reduce information leakage by restricting the mutual information that Eve can obtain. Both RM (2,2) and (3,3) codes maintain a mutual information level below that of uncoded messages, for a certain range, underscoring the security benefits of using RM coding for Physical-layer security.

\begin{figure}[!ht]
\centering
\includegraphics[width=1\linewidth]{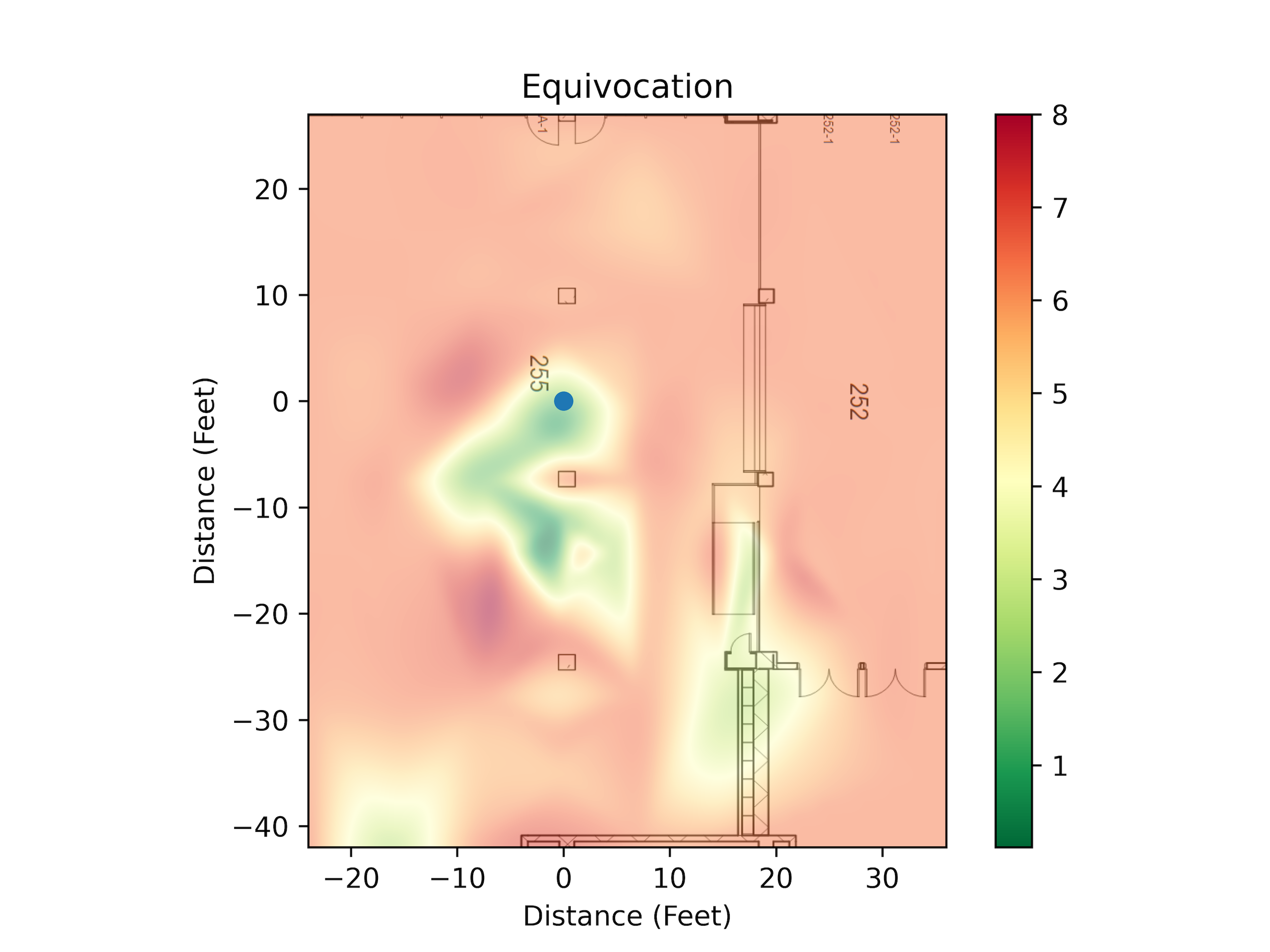}
\caption{Heat map of equivocation.}
\label{fig:equivocation_heatmap}
\end{figure}

Figure~\ref{fig:equivocation_heatmap} presents a heatmap of equivocation levels in an indoor environment, with the transmitter (Alice) represented by the blue dot. The x- and y-axes denote the distance in feet from Alice, while the color gradient illustrates equivocation across various locations. Green areas correspond to low equivocation, where information leakage is more likely, while red areas represent high equivocation, indicating greater uncertainty and stronger security.

The heatmap reveals that the areas closest to Alice, shown in green, exhibit the lowest levels of equivocation. This suggests that an eavesdropper positioned near Alice would experience less uncertainty about the transmitted message, making these locations more vulnerable to information leakage. This pattern is consistent with Physical-layer security expectations, where close proximity to the transmitter increases the eavesdropper’s ability to decode the message accurately.

As the distance from Alice increases, the equivocation levels also increase, with red regions indicating zones of higher equivocation. These red areas, further from Alice, suggest that an eavesdropper would experience significant uncertainty about the transmitted data, enhancing security in these regions. The spatial distribution of equivocation illustrates that security improves as distance from Alice increases, and environmental factors such as walls and obstacles likely contribute to attenuating the signal further, thereby increasing equivocation.

\begin{figure}[!ht]
\centering
\includegraphics[width=1\linewidth]{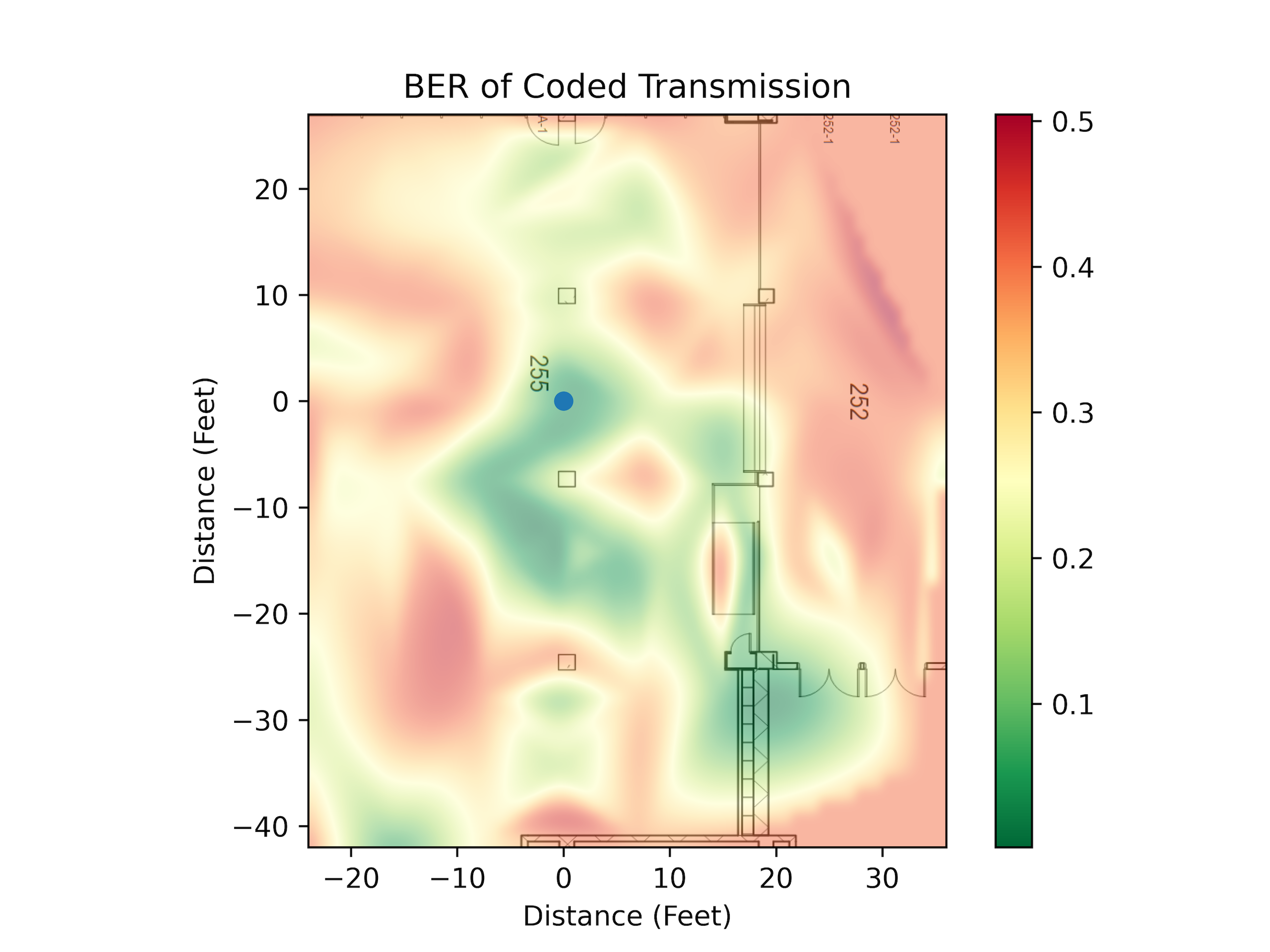}
\caption{Heat map of BER for RM(4,4) transmission.}
\label{fig:ber_coded}
\end{figure}

\begin{figure}[!ht]
\centering
\includegraphics[width=1\linewidth]{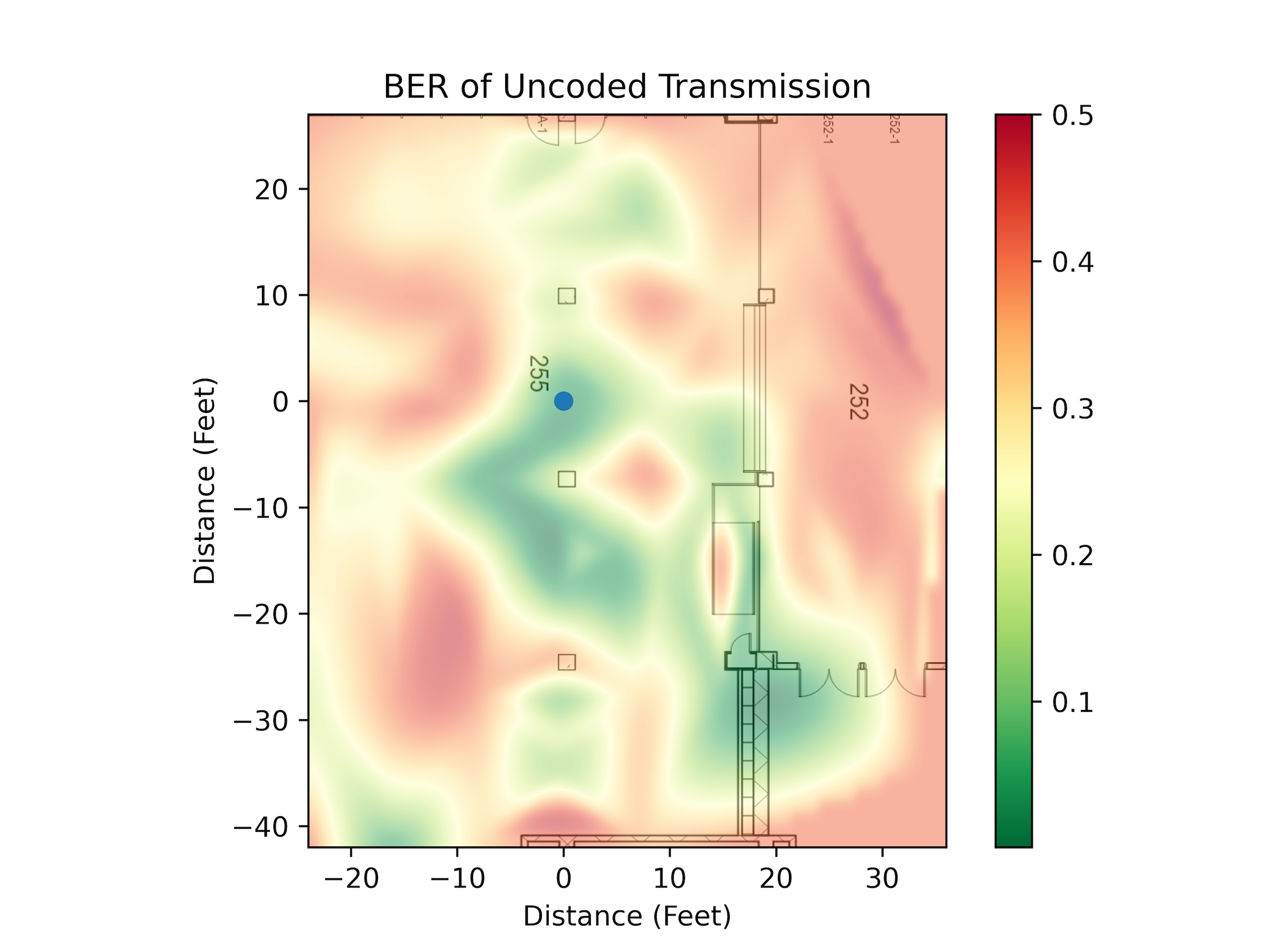}
\caption{Heat map of BER for uncoded transmission.}
\label{fig:ber_uncoded}
\end{figure}

\begin{figure}[!ht]
\centering
\includegraphics[width=1\linewidth]{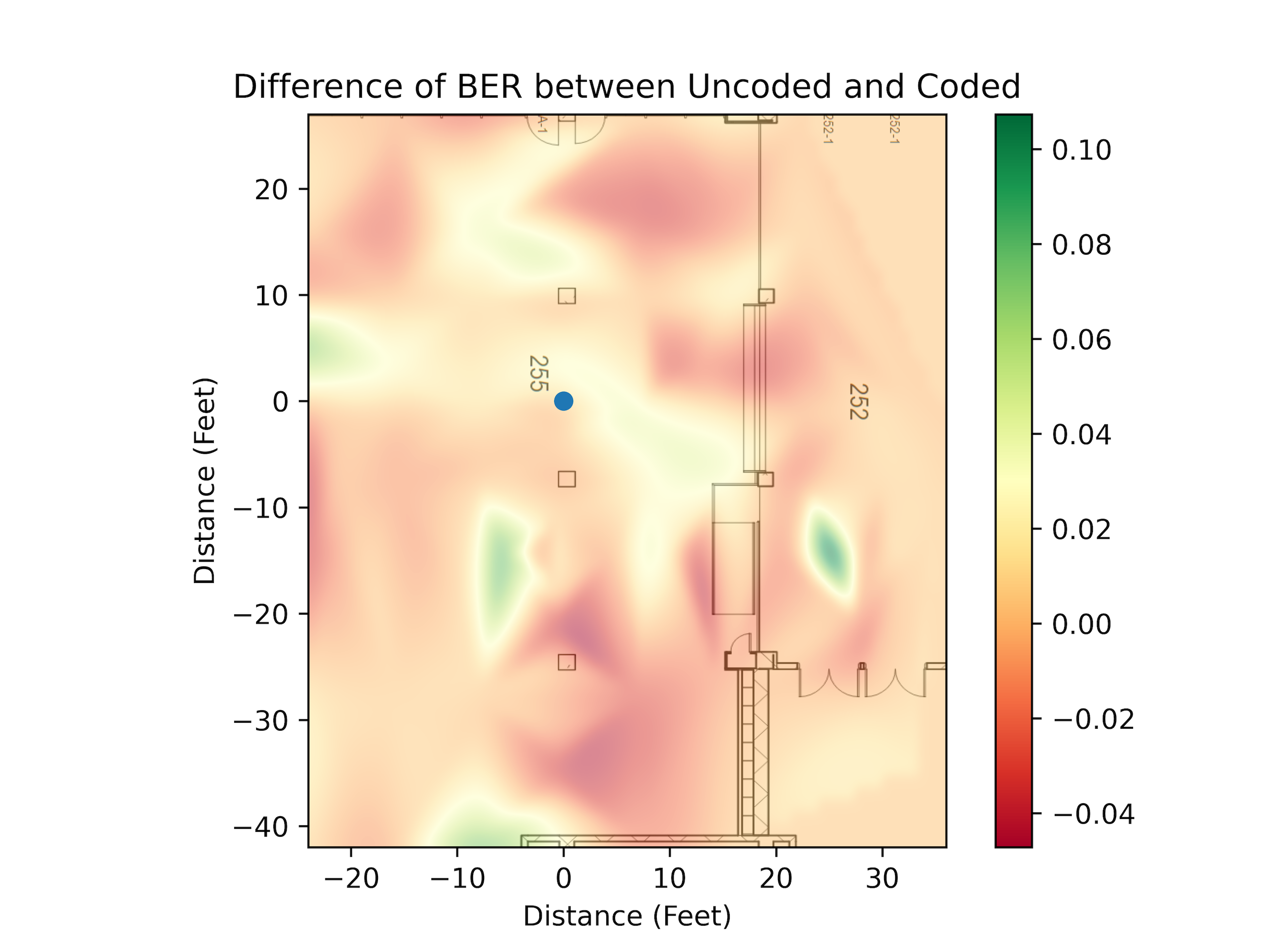}
\caption{Heat map of BER for difference between coded and uncoded transmission.}
\label{fig:ber_difference}
\end{figure}

The BER heatmaps in Figures~\ref{fig:ber_coded}, \ref{fig:ber_uncoded}, and \ref{fig:ber_difference} compare coded and uncoded transmissions in an indoor environment, with Alice’s location marked by the blue dot. Figures~\ref{fig:ber_coded} and \ref{fig:ber_uncoded} show that both coded and uncoded transmissions exhibit similar BER distributions, with red regions near Alice indicating lower BER and green regions farther away showing higher BER due to signal attenuation.

Figure~\ref{fig:ber_difference}, which presents the difference in BER between uncoded and coded transmissions, reinforces this similarity by displaying minimal differences across the environment. This indicates that, in terms of BER alone, the coded transmission does not provide a significant advantage over the uncoded one when compared to equivocation map of the area.


However, despite the similar BER performance, the coded transmission offers a distinct Physical-layer security advantage, as demonstrated in the equivocation heatmap (Figure~\ref{fig:equivocation_heatmap}). While BER is nearly the same for both coded and uncoded cases, the equivocation heatmap reveals that the coded transmission introduces substantial uncertainty for an eavesdropper, creating areas of high equivocation (red) that effectively protect the transmitted information. This added layer of security highlights the effectiveness of the coded transmission in limiting information leakage, which is not apparent from BER analysis alone.

\section{C\normalfont onclusion}
This study demonstrates the practical application of wiretap coding for Physical-layer security using Reed-Muller (RM) codes implemented on software-defined radios (SDRs) in a real indoor environment. The structured coset coding approach, leveraging the hierarchical properties of RM codes, provides a secure communication framework that allows a legitimate receiver (Bob) to decode messages with low bit error rates (BER) while maintaining a high BER for an eavesdropper (Eve).

By analyzing the bit error rate across varying distances and employing Mutual Information Neural Estimation (MINE) to calculate the leakage rate, the results validate the effectiveness of coset codes in achieving security. The findings highlight that, even in practical environments with real-world channel effects, coset codes can effectively create secrecy regions where Bob’s channel quality surpasses that of Eve, thus ensuring confidentiality.

The experimental results also underscore the potential of RM codes for secure wireless communications, particularly in low-latency and short blocklength scenarios. The adaptability of RM codes in balancing reliability and security through coset-based randomization suggests they are a strong candidate for Physical-layer security applications in modern communication systems.

Future research could explore the use of other coding schemes in similar setups and investigate the impact of different environmental conditions. Further refinement of the MINE framework may also provide more precise measurements of information leakage, enhancing the understanding of RM codes’ capabilities in securing communications against advanced eavesdropping threats.

\bibliographystyle{IEEEtran}
\bibliography{IEEEabrv,mybib}

\end{document}